\documentclass[aps,pra,preprint,superscriptaddress]{revtex4}

\usepackage{amsmath}
\usepackage{mathrsfs}
\usepackage{comment}
\usepackage{amssymb}
\usepackage{graphicx}
\usepackage{subfigure}
\usepackage[colorlinks,
            linkcolor=blue,
            anchorcolor=blue,
            citecolor=blue]{hyperref}
\newtheorem{theorem}{Theorem}
\newtheorem{corollary}{Corollary}

\def\ra{\rangle}
\def\la{\langle}

\allowdisplaybreaks[1]

\begin{document}

\title{Wigner-Yanase skew information-based uncertainty relations for quantum channels}

\author{Qing-Hua Zhang}
\email[]{qhzhang@csust.edu.cn}

%\thanks{}
\affiliation{School of Mathematics and Statistics, Changsha University of Science and Technology, Changsha 410114, China}

\author{Shao-Ming Fei}
\email[]{feishm@cnu.edu.cn}
\affiliation{School of Mathematical Sciences, Capital Normal University,
Beijing 100048, China}
\affiliation{Max-Planck-Institute for Mathematics in the Sciences, 04103 Leipzig, Germany}

%\date{\today}

\begin{abstract}
The Wigner-Yanase skew information stands for the uncertainty about the information on the values of observables not commuting with the conserved quantity. The Wigner-Yanase skew information-based uncertainty relations can be regarded as a complementarity to the conceptual Heisenberg uncertainty principle. We present tight uncertainty relations in both product and summation forms for two quantum channels based on the Wigner-Yanase skew information. We show that our uncertainty inequalities are tighter than the existing ones.
\end{abstract}

\maketitle

\section{Introduction}

A fundamental characteristic of quantum mechanics is the intrinsic uncertainty associated with measuring incompatible observables, named as Heisenberg uncertainty principle~\cite{heisenberg1927uber}. For arbitrary two observables $M_1$ and $M_2$, the famous Heisenberg-Robertson uncertainty relation exhibits the limitation on the precision of simultaneously measuring the two observables ~\cite{robertson1929the}, $\Delta M_1\Delta M_2\geq \frac{1}{2}|\la \psi |[M_1,M_2]|\psi\ra|$,
where $\Delta \Omega=\sqrt{\la \Omega^2\ra-\la\Omega\ra^2}$ is the standard deviation of an observable $\Omega$ with respect to the measured state $|\psi\ra$ and $[M_1,M_2]=M_1M_2-M_2M_1$. Deutsch~\cite{deutsch1983uncertainty} introduced an uncertainty relation based on the Shannon entropy to characterize the measurement incompatibility of observables, $H(M_1)+H(M_2)\geq 2\log_2(\frac{2}{1+\sqrt{c}})$, where $H(M_1)$ is the Shannon entropy given by $H(M_1)=-\sum_i p_i\log_2 p_i$ with $p_i=\langle \psi_i|\rho|\psi_i\rangle$, $H(M_2)=-\sum_i q_i\log_2 q_i$ with $q_i=\langle \phi_i|\rho|\phi_i\rangle$, $c=\max_{jk}|\langle \psi_j|\phi_k\rangle|^2$, $|\psi_j\ra$ and $|\phi_k\ra$ are respectively the eigenvectors of the observables $M_1$ and $M_2$.

Riding the waves of information theory, a variety of characterizations and quantifications of uncertainty relations have been established, including but not limited to the ones in terms of variance~\cite{kraus1987complementary,maassen1988generalized,PhysRev.34.163,schrodinger1930sitzungsberichte,PhysRevLett.113.260401,Kennard1927Zur,mondal2017tighter,PhysRevResearch.4.013076,PhysRevResearch.4.013075,wu2023parameterized}, entropy~\cite{Maassen1988PhysRevLett.60.1103,Wu2009PhysRevA.79.022104,Coles2017RevModPhys.89.015002,wu2022tighter}, noise and disturbance~\cite{buschPhysRevLett.111.160405}, successive measurement~\cite{DeutschPhysRevLett.50.631,DistlerPhysRevA.87.062112}, majorization technique~\cite{Pucha_a_2013,friedland2013universal}, skew information~\cite{luo2003wigner,zhang2021a,zhang2021tighter,ma2022product,zhang2023note}, etc.. These uncertainty relations play an important role in a wide range of applications in such as quantum gravity~\cite{hall2005exact}, quantum cryptography~\cite{Fuchs1996Quantum,berta2010uncertainty}, entanglement detection~\cite{PhysRevLett.92.117903,zhang2021multipartite}, nonlocality~\cite{doi:10.1126/science.1192065}, quantum steering~\cite{SchneelochPhysRevA.87.062103} and quantum metrology~\cite{PhysRevLett.96.010401}.

The Wigner-Yanase skew information of a quantum state $\rho$ with respect to an observable $A$ is given by~\cite{wigner1963information,luo2018coherence,zhang2021a},
$I_{\rho}(A)=\frac{1}{2}{\mathbf{tr}}([\sqrt{\rho},A]^\dagger[\sqrt{\rho},A])$.
Wigner and Yanase proved that this quantity satisfies all the desirable requirements of an information measure. Luo ~\cite{luo2003wigner} supplied the skew information perspective to quantify the Heisenberg uncertainty principle and Bohr's complementarity principle.

As the most general description of quantum measurement, quantum channel plays a pivotal role in quantum information processing \cite{nielsen2002quantum}. For any quantum channel $\Phi$ with Kraus representation, $\Phi(\rho)=\sum_{i=1}^{n}K_i\rho K_i^\dagger$, Luo defined in Ref.~\cite{luo2018coherence} the Wigner-Yanase skew information of the channel $\Phi$, \begin{equation}
I_{\rho}(\Phi)=\sum_{i=1}^{n}I_{\rho}(K_i),
\end{equation}
where $I_{\rho}(K_i)=\frac{1}{2}\mathbf{tr}([\sqrt{\rho},K_i]^\dagger[\sqrt{\rho},K_i])$. The quantity $I_{\rho}(\Phi)$ is well-defined as it is independent of the choice of the Kraus representation. Moreover, the quantity $I_{\rho}(\Phi)$ can be regarded as a bona fide measure for coherence as well as the quantum uncertainty of $\rho$ with respect to the quantum channel $\Phi$.

Naturally, Fu $et\ al.$ established the sum-form uncertainty relation via Wigner-Yanase skew information for two quantum channels $\Psi$ and $\Phi$~\cite{fu2019skew},
\begin{equation}\label{fu2019ur}
I_{\rho}(\Psi)+I_{\rho}(\Phi) \geq \max_{\pi \in S_n}\frac{1}{2}\sum_{i=1}^n I_{\rho}(L_i\pm K_{\pi(i)}),
\end{equation}
where $\Psi=\sum_{i=1}^nL_i \rho L_i^\dagger$, $\Phi=\sum_{i=1}^nK_i \rho K_i^\dagger$, and $\pi \in S_n$ is an arbitrary $n$-element permutation. Very recently, Zhou $et\ al.$ established the product-form uncertainty relation for two channels~\cite{zhou2023uncertainty},
\begin{equation}\label{zhou2023ur}
I_{\rho}(\Psi)I_{\rho}(\Phi) \geq \frac{1}{4}\sum_{i=1}^n\sum_{j=1}^n |{\mathbf{tr}}([\sqrt{\rho},L_i]^\dagger[\sqrt{\rho},K_j])|^2.
\end{equation}

In this paper, we focus on improving the lower bounds of uncertainty relations based on Wigner-Yanase skew information in product-form and sum-form for two quantum channels. The lower bounds of our uncertainty inequalities are tighter than the existing ones~\cite{fu2019skew,zhou2023uncertainty}. Detailed examples are presented to illustrate the advantages of our results. In addition, we also discuss the uncertainty relations for unitary channels.

\section{Uncertainty relations in product form}
In this section, we study tighter product-form uncertainty relations based on the vectorization of commutators. We consider $d$-dimensional quantum system whose Hilbert space $H_d$ is spanned by the set of computational basis $|i\ra,\ i=1,2,\dots,d$. Let $G=(g_{ij})_{l\times p}$ be a rectangular matrix with entries $g_{ij}$. The vectorization of $G$ is given by the vector $|G\rangle=(g_{11},\dots,g_{l1},\dots,g_{1p},\dots,g_{lp})^T$, where $T$ denotes the transpose. It is verified that $|GM\ra=(I\otimes G)|M\ra$ for any matrix $M$ and identity $I$ in suitable size.

Define the correlation measure of observables $A$ and $B$ via Wigner-Yanase skew information,
\begin{equation}
Corr_{\rho}^{WY}(A,B)=\frac{1}{2} \mathbf{tr}([\sqrt{\rho},A]^\dagger[\sqrt{\rho},B])=\frac{1}{2} \la \tilde{A}|\tilde{B}\ra,
\end{equation}
where $\tilde{A}=[\sqrt{\rho},A]$, $\tilde{B}=[\sqrt{\rho},B]$ and $\la \cdot|\cdot\ra$ denotes the inner product in Hilbert space. Obviously, $I_{\rho}(A)=Corr_{\rho}^{WY}(A,A)$.

Let $\Phi$ be a quantum channel with Kraus representation $\Phi(\rho)=\sum_{i=1}^{n}K_i\rho K_i^\dagger$. The Wigner-Yanase skew information of the channel with respect to $\rho$ can be rewritten as
\begin{equation}
I_{\rho}(\Phi)=\sum_{i=1}^{n}I_{\rho}(K_i)=\sum_{i=1}^{n} Corr_{\rho}^{WY}(K_i,K_i).
\end{equation}
Let $X=(X_{st})_{d\times d}$ and $Y=(Y_{st})_{d\times d}$ be any complex matrices. According to the Cauthy-Schwarz inequality, the following inequality holds
\begin{equation}\label{cauthy_schwarz}
\begin{aligned}
\la X|X\ra\la Y|Y\ra=\sum_{s,t}^{d} |x_{st}|^2\sum_{s^\prime,t^\prime}^{d}|y_{s^\prime t^\prime}|^2\geq (\sum_{s,t}^{d} |x_{s,t}^*y_{s,t}|)^2\geq |\sum_{s,t}^{d} x_{s,t}^*y_{s,t}|^2=|\la X|Y\ra|^2.
\end{aligned}
\end{equation}
We have the following theorem according to the inequality sequence.

\begin{theorem}\label{wysi_ur_c_th1}
Let $\Psi(\rho)=\sum_{i=1}^{n}L_i\rho L_i^\dagger$ and $\Phi(\rho)=\sum_{i=1}^{n}K_i\rho K_i^\dagger$ be two quantum channels. 
We have the following tighter product-form uncertainty relation,
\begin{equation}
I_{\rho}(\Psi)I_{\rho}(\Phi)\geq \frac{1}{4}\sum_{i,j}^n\sum_{s,t}^d |\la ^{s}\tilde{L}_i|\tilde{K}_j^t \ra |^2,
\end{equation}
where $^{s}\tilde{L}_i=|s\ra\la s|[\sqrt{\rho},L_i]$ and $\tilde{K}_j^t =[\sqrt{\rho},K_j]|t\ra\la t|$.
\end{theorem}

\noindent{\textit{Proof.}} By employing the inequality sequence (\ref{cauthy_schwarz}), we have
\begin{equation}
\begin{aligned}
I_{\rho}(L_i)I_{\rho}(K_j)&= \frac{1}{4}\la\tilde{L}_i|\tilde{L}_i\ra\la\tilde{K}_j|\tilde{K}_j\ra\\
&=\frac{1}{4}\sum_{s,t}^{d}| \la t |\tilde{L}_i^\dagger|s \ra |^2\sum_{s^\prime,t^\prime}^{d}| \la s^\prime |\tilde{K}_j|t^\prime \ra |^2\\
&\geq \frac{1}{4} (\sum_{s,t}^{d}| \la t |\tilde{L}_i^\dagger|s \ra \la s |\tilde{K}_j|t \ra |)^2\\
&=\frac{1}{4}(\sum_{s,t}^d |\la ^{s}\tilde{L}_i|\tilde{K}_j^t \ra |)^2.
\end{aligned}
\end{equation}
According to the definition of Wigner-Yanase skew information for quantum channels we obtain
\begin{equation}
I_{\rho}(\Psi)I_{\rho}(\Phi)=\sum_{i,j}^n I_{\rho}(L_i)I_{\rho}(K_j)\geq \frac{1}{4}\sum_{i,j}^n(\sum_{s,t}^d |\la ^{s}\tilde{L}_i|\tilde{K}_j^t \ra |)^2.
\end{equation}
The proof is completed. $\Box$

We remark that from the inequality sequence (\ref{cauthy_schwarz}), it is obvious that our lower bound in Theorem \ref{wysi_ur_c_th1} is more tighter than that of the uncertainty relation (\ref{zhou2023ur}).

Furthermore, we notice that the order of vectorization of $|\tilde{L_i}\ra$ or $|\tilde{K_j}\ra$ has significant impact on the lower bound in Theorem \ref{wysi_ur_c_th1}. In Ref.~\cite{xiao2022near}, Xiao $et. al$ introduced several `near optimal' bounds for incompatible observables in term of quantum variances. Let $X^\downarrow$ be a rearranged matrix of $X$, whose vectorization is rearranged in non-increasing order, $|\la  s|X^\downarrow|t\ra|\geq |\la s+1|X^\downarrow|t\ra|\geq |\la s^\prime |X^\downarrow|{t+1}\ra|$ for any $s$, $s^\prime$ and $t$. Let $\pi$ be any permutation belonging to the $d^2$-elements permutation group $S_{d^2}$, and $X^\pi$ the rearranged matrix of $X$ under permutation $\pi$. We have
\begin{equation}
I_{\rho}(L_i)I_{\rho}(K_j)= \frac{1}{4}\la\tilde{L}_i|\tilde{L}_i\ra\la\tilde{K}_j^\downarrow|\tilde{K}_j^\downarrow \ra=\frac{1}{4}\la\tilde{L}_i |\tilde{L}_i\ra\la\tilde{K}_j^\pi|\tilde{K}_j^\pi \ra.
\end{equation}
Rewriting $|X^\downarrow\ra$ to be $|X^\downarrow\ra=(x_1^\prime,x_2^\prime,\dots,x_{d^2}^\prime)$ such that $x_i^\prime\geq x_{i+1}$, and so does $|Y^\downarrow\ra$. Since the rearranged inequality relation always holds under any permutation $\pi\in S_{d^2}$ acting on all elements of $|Y\ra$,
\begin{equation}\label{rearrange_ineq}
\sum_i^{d^2}|x_i^\prime y_i^\prime|\geq \sum_i^{d^2}|x_i y_{\pi (i)}|,
\end{equation}
we obtain the following theorem.

\begin{theorem}\label{wysi_ur_c_th2}
The following tighter uncertainty relation based on Wigner-Yanase skew information for quantum channels holds,
\begin{equation}
I_{\rho}(\Psi)I_{\rho}(\Phi)\geq  \frac{1}{4}\sum_{i,j}^n(\sum_{s,t}^d |\la ^{s}{\tilde{L}_i^\downarrow} |{\tilde{K}_j}^{\downarrow t} \ra |)^2.
\end{equation}
\end{theorem}

Unitary channels are also used a lot in quantum computation and quantum information processing~\cite{nielsen2002quantum}. For an arbitrary unitary channel $U(\rho)=U\rho U^\dagger$, the Wigner-Yanase skew information of $\rho$ with respect to the channel is given by
$I_{\rho}(U)=\frac{1}{2}\mathbf{tr} ([\sqrt{\rho},U]^\dagger[\sqrt{\rho},U])$.
Next we consider the skew information-based uncertainty relation for arbitrary two unitary channels $U$ and $V$. Directly from Theorom \ref{wysi_ur_c_th1}, we have the following uncertainty relation:

\begin{corollary}
Let $U$ and $V$ be any two unitary channels with $U(\rho)=U^\dagger \rho U$ and $V(\rho)=V^\dagger \rho V$, respectively. The following tighter uncertainty relation holds,
\begin{equation}
I_{\rho}(U)I_{\rho}(V)\geq \frac{1}{4}(\sum_{s,t}^d |\la ^{s}{\tilde{U}^\downarrow} |{\tilde{V}}^{\downarrow t} \ra |)^2.
\end{equation}
\end{corollary}

Besides the product-form uncertainty relations, the sum-form uncertainty relations are considered to be useful in complementary to the product-form uncertainty~\cite{robertson1929the}. We study the sum-form uncertainty relations for quantum channels in the following. From the parallelogram law, the summation of Wigner-Yanase skew information for two Kraus operators $L_i$ and $K_j$ can be decomposed into
\begin{equation}
I_{\rho}(L_i)+I_{\rho}(K_j)=\frac{1}{2}I_{\rho}(L_i+K_j)+\frac{1}{2}I_{\rho}(L_i-K_j).
\end{equation}
Combining with the rearrangement inequality (\ref{rearrange_ineq}), we obtain the following result.

\begin{theorem}\label{wysi_ur_c_th3}
Let $\Psi$ and $\Phi$ be two channels with Kraus decomposition $\Psi(\rho)=\sum_{i=1}^{n}L_i\rho L_i^\dagger$ and $\Phi(\rho)=\sum_{i=1}^{n}K_i\rho K_i^\dagger$. The following sum-form uncertainty relation holds,
\begin{equation}
\begin{aligned}
I_{\rho}(\Psi)+I_{\rho}(\Phi)\geq &\frac{1}{4}\sum_{i,j}\sum_{s,t}^d |\la {^{s}\tilde{L}_i+^{s}\tilde{K}_j}|(\tilde{L}_i^{\pi_i^1})^t+(\tilde{K}_j^{\pi_i^1})^t\ra|\\
&+\frac{1}{4}\sum_{i,j}\sum_{s,t}^d |\la {^{s}\tilde{L}_i-^{s}\tilde{K}_j}|(\tilde{L}_i^{\pi^2_j})^t-(\tilde{K}_j^{\pi^2_j})^t\ra|,
\end{aligned}
\end{equation}
where $\pi_i^1,\pi^2_j\in S_{d^2}$.
\end{theorem}

{\textit{Proof.}} For any two Kraus operators $L_i$ and $K_j$, one has
\begin{equation}
\begin{aligned}
I_{\rho}(L_i)+I_{\rho}(K_j)&=\frac{1}{2}I_{\rho}(L_i+K_j)+\frac{1}{2}I_{\rho}(L_i-K_j)\\
&=\frac{1}{4}\la \tilde{L}_i+\tilde{K}_j|\tilde{L}_i+\tilde{K}_j\ra+\frac{1}{4}\la \tilde{L}_i-\tilde{K}_j|\tilde{L}_i-\tilde{K}_j\ra\\
&=\sum_{s,t}\frac{1}{4}|\la s|\tilde{L}_i+\tilde{K}_j|t\ra|^2+\sum_{s^\prime,t^\prime}\frac{1}{4}|\la s^\prime|\tilde{L}_i-\tilde{K}_j|t^\prime\ra|^2\\
&\geq \sum_{s,t}\frac{1}{4}|\la t|\tilde{L}^{\dagger}_i+\tilde{K}^{\dagger}_j|s\ra\la s|\tilde{L}_i^{\pi^1_i}+\tilde{K}_j^{\pi^1_i}|t\ra|\\
&+\sum_{s,t}\frac{1}{4}|\la t|\tilde{L}^{\dagger}_i-\tilde{K}^{\dagger}_j|s\ra\la s|\tilde{L}_i^{\pi^2_j}-\tilde{K}_j^{\pi^2_j}|t\ra|\\
&=\frac{1}{4}\sum_{s,t}^d |\la {^{s}\tilde{L}_i+^{s}\tilde{K}_j}|(\tilde{L}_i^{\pi^1_i})^t+(\tilde{K}_j^{\pi^1_i})^t\ra|\\
&+\frac{1}{4}\sum_{s,t}^d |\la {^{s}\tilde{L}_i-^{s}\tilde{K}_j}|(\tilde{L}_i^{\pi^2_j})^t-(\tilde{K}_j^{\pi^2_j})^t\ra|.
\end{aligned}
\end{equation}
Thus for any two quantum channels we get
\begin{equation}
\begin{aligned}
I_{\rho}(\Psi)+I_{\rho}(\Phi)&=\sum_{i,j}\frac{1}{2}I_{\rho}(L_i+K_j)+\frac{1}{2}I_{\rho}(L_i-K_j)\\
& \geq \frac{1}{4}\sum_{i,j}\sum_{s,t}^d |\la {^{s}\tilde{L}_i+^{s}\tilde{K}_j}|(\tilde{L}_i^{\pi^1_i})^t+(\tilde{K}_j^{\pi^1_i})^t\ra|\\
&+ \frac{1}{4}\sum_{i,j}\sum_{s,t}^d |\la {^{s}\tilde{L}_i-^{s}\tilde{K}_j}|(\tilde{L}_i^{\pi^2_j})^t-(\tilde{K}_j^{\pi^2_j})^t\ra|.
\end{aligned}
\end{equation}
$\Box$

In particular, for unitary channels we have

\begin{corollary}
Let $U$ and $V$ be any two unitary channels with $U(\rho)=U^\dagger \rho U$ and $V(\rho)=V^\dagger \rho V$, respectively. 
The following near-optimal uncertainty relation holds,
\begin{equation}
\begin{aligned}
I_{\rho}(U)+I_{\rho}(V)
\geq  \frac{1}{4}\sum_{s,t}^d |\la {(^{s}\tilde{U}+^{s}\tilde{V})}|(\tilde{U}^{\pi^1})^t+(\tilde{V}^{\pi^1})^t\ra|+ \frac{1}{4}\sum_{s,t}^d |\la {(^{s}\tilde{U}-^{s}\tilde{V})}|(\tilde{U}^{\pi^2})^t-(\tilde{V}^{\pi^2})^t\ra|.
\end{aligned}
\end{equation}
\end{corollary}

\begin{figure}[htbp]
 \centering
 \subfigure[]
 {
 \label{subfig:a} %% label for first subfigure
 \includegraphics[width=7.8cm]{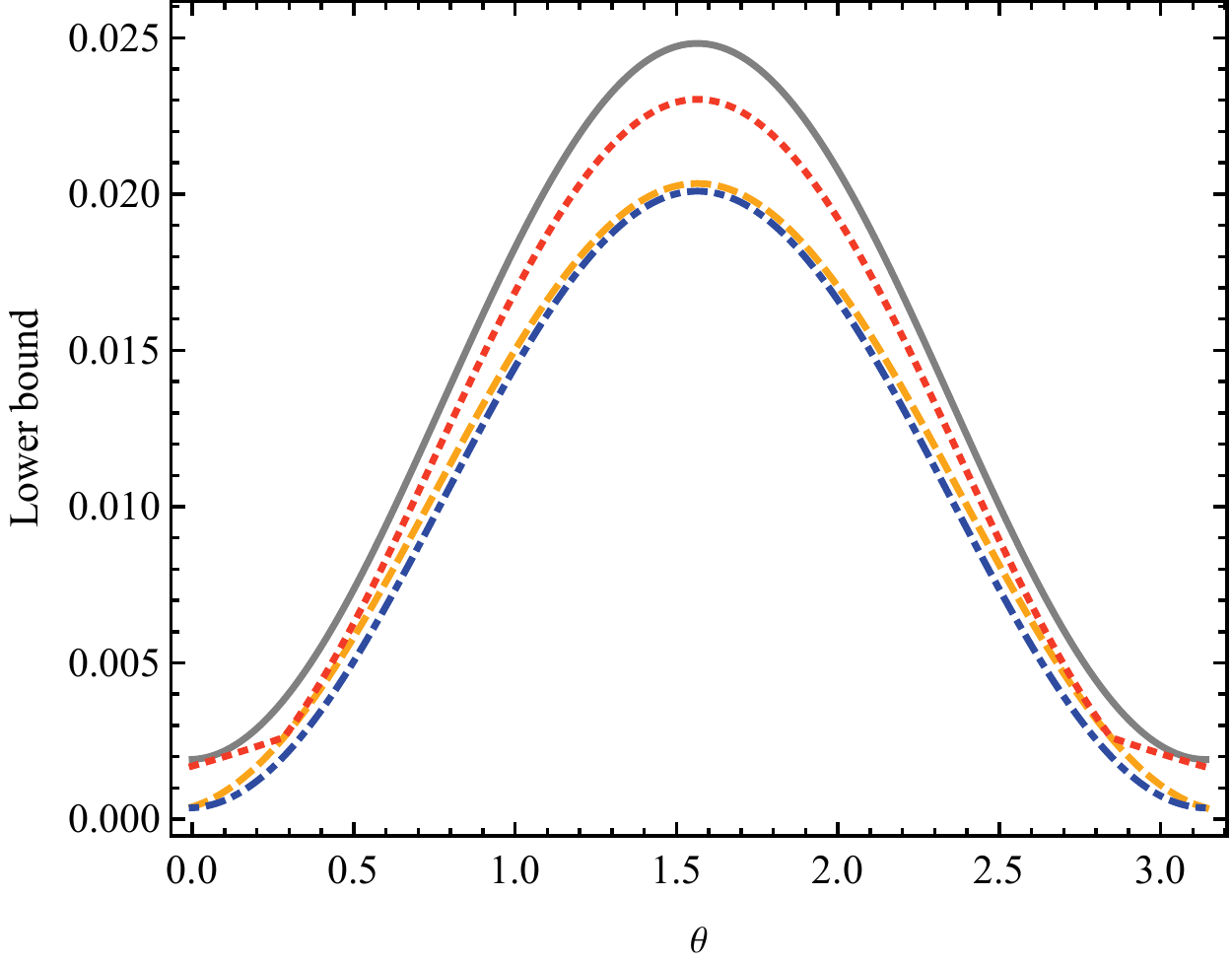}
 }
 \subfigure[]
 {
 \label{subfig:b} %% label for second subfigure
 \includegraphics[width=7.8cm]{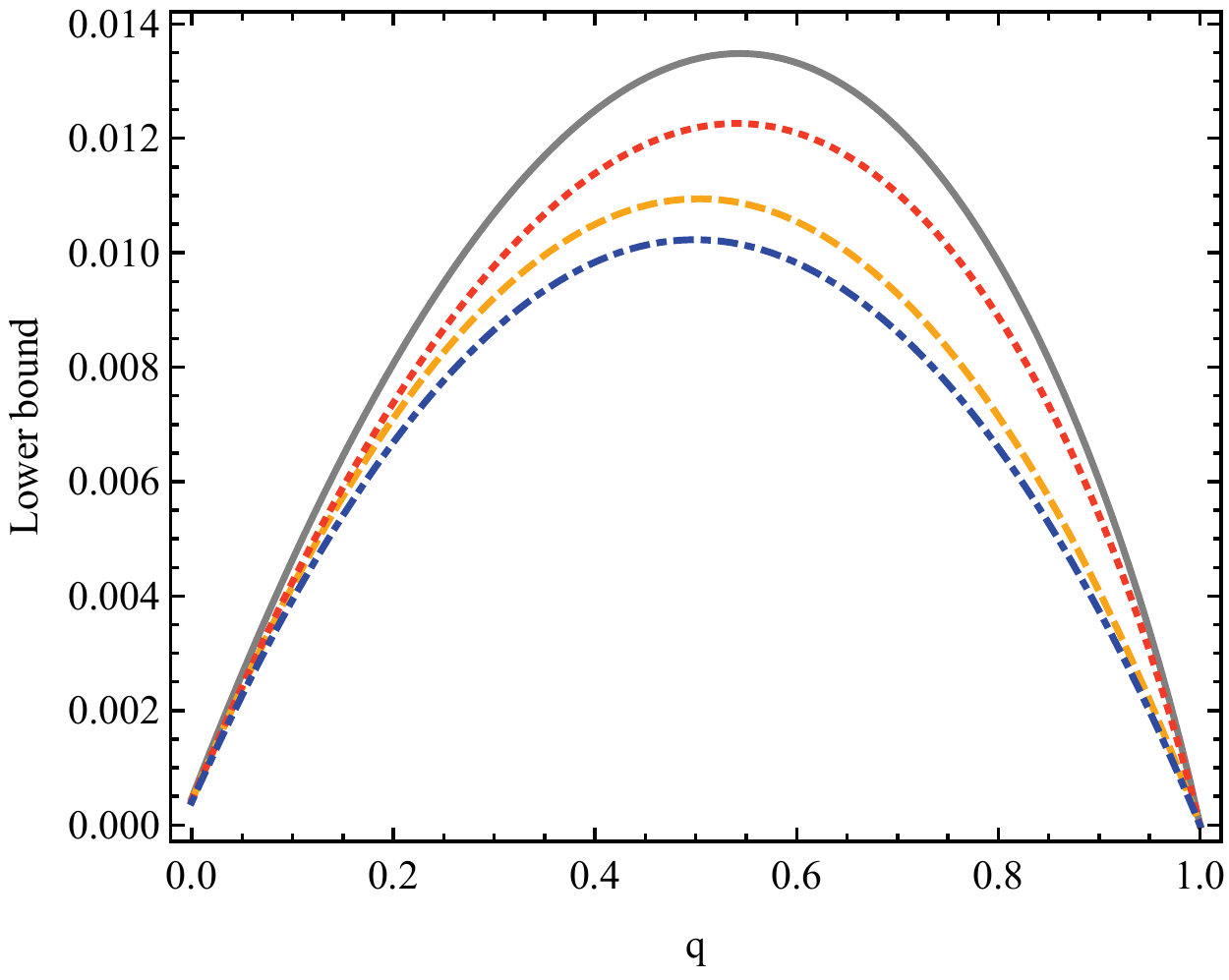}}
  \subfigure[]
 {
 \label{subfig:c} %% label for second subfigure
 \includegraphics[width=7.8cm]{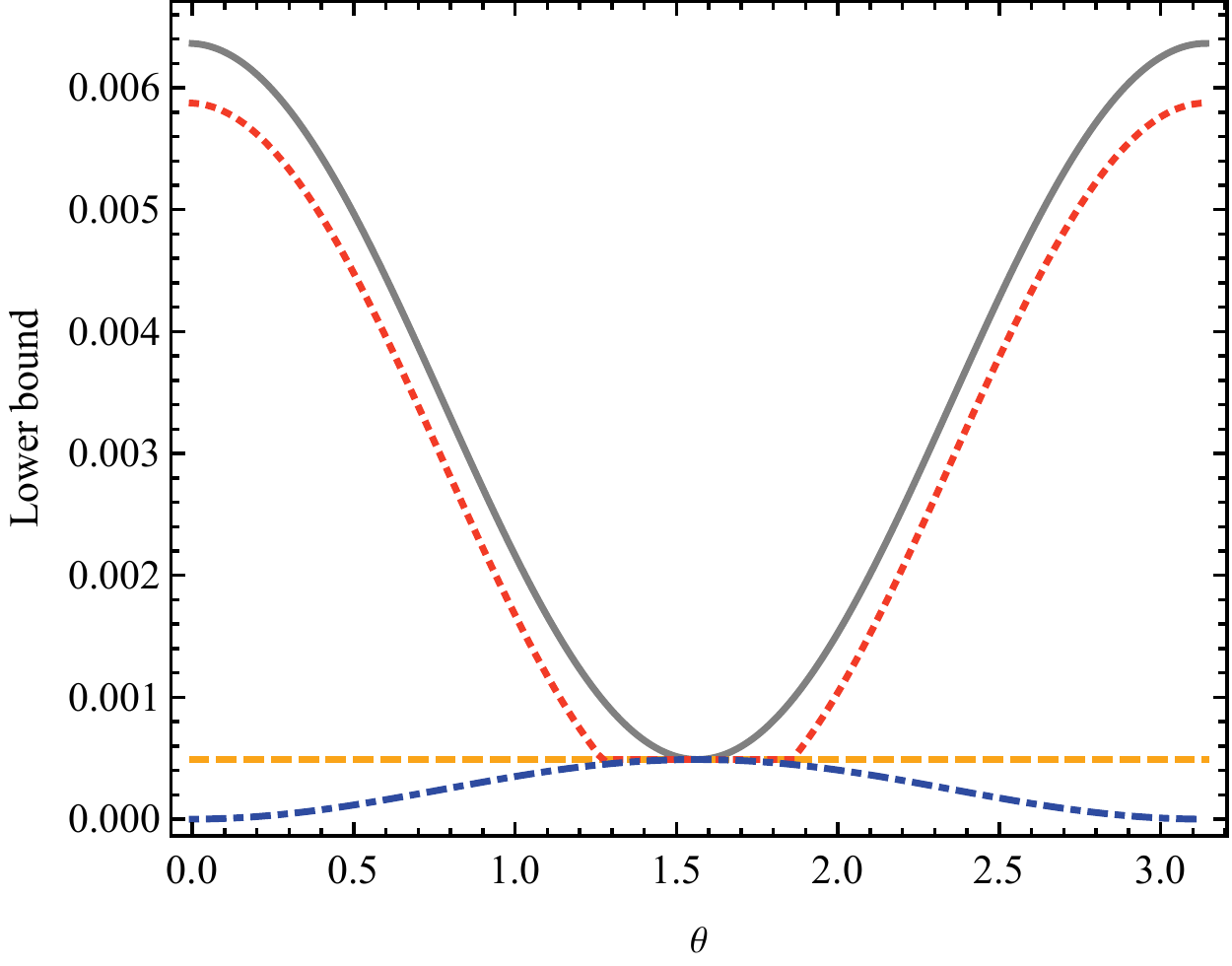}}
  \subfigure[]
 {
 \label{subfig:d} %% label for second subfigure
 \includegraphics[width=7.8cm]{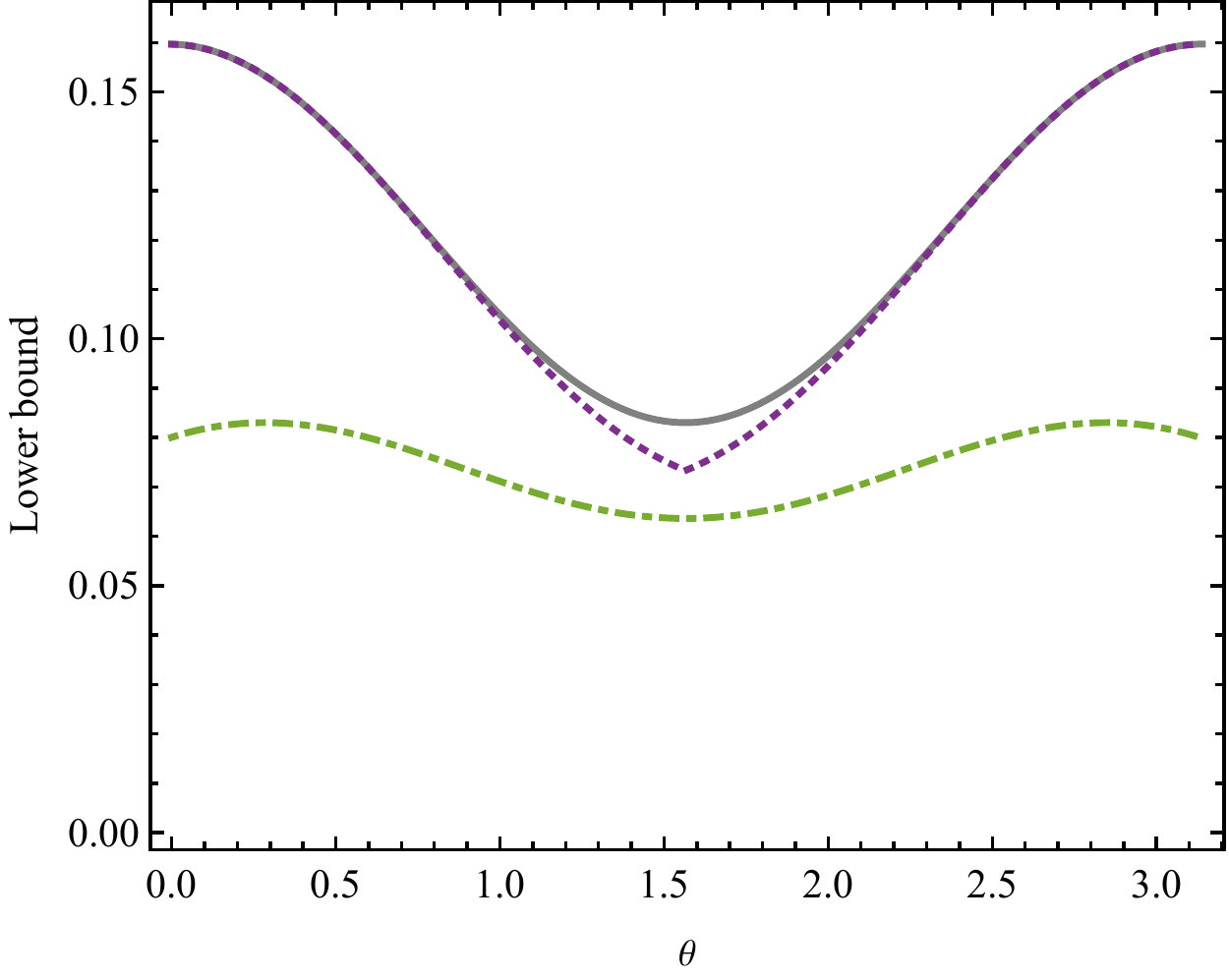}}
 \caption{The grey (solid) curve represents the product-form or sum-form uncertainties with respect to the state $\rho$. The yellow (dashed) curve, the red (dotted) curve and the purple (dashed) represent respectively the lower bounds in our Theorem \ref{wysi_ur_c_th1}, \ref{wysi_ur_c_th2} and \ref{wysi_ur_c_th3}. The blue (dot-dashed) curve and the green (dot-dashed) represent the lower bounds of Zhou's result (\ref{zhou2023ur}) and Fu's result (\ref{fu2019ur}), respectively. (a) The comparison among our Theorem \ref{wysi_ur_c_th1}, \ref{wysi_ur_c_th2} and Zhou's result (\ref{zhou2023ur}) for the phase damping channel $\phi$ and the bit flip channel $\Lambda$ with $q=0.5$. (b) The comparison among the Theorem \ref{wysi_ur_c_th1}, \ref{wysi_ur_c_th2} and Zhou's result (\ref{zhou2023ur}) for the phase damping channel $\phi$ and the bit flip channel $\Lambda$ with $\theta=\pi/4$. (c) The comparison among the Theorem \ref{wysi_ur_c_th1}, \ref{wysi_ur_c_th2} and Zhou's result (\ref{zhou2023ur}) for the unitary operators $U$ and $V$. (d) The comparison between the Theorem \ref{wysi_ur_c_th3} and Fu's result (\ref{fu2019ur}) for the unitary operators $U$ and $V$.}
 \label{fig}
 \end{figure}

We take an example to illustrate the performance of these uncertainty relations. Let us consider the mixed state $\rho=\frac{1}{2}(I+\vec{r}\cdot\vec{\sigma})$ with $\vec{r}=(\frac{\sqrt{3}}{2}\cos\theta,\frac{\sqrt{3}}{2}\sin\theta,\frac{1}{4})$, where $\sigma_x$,$\sigma_y$,$\sigma_z$ are Pauli matrices. To illustrate that our results are better than the ones in Ref.~\cite{fu2019skew} and Ref.~\cite{zhou2023uncertainty}, we firstly consider two quantum channels: the phase damping channel $\phi(\rho)=\sum_{i=1}^2L_i\rho (L_i)^\dagger$ with $L_1=|0\ra\la0|+\sqrt{1-q}|1\ra\la1|$ and $L_2=\sqrt{q}|1\ra\la1|$, and the bit flip channel $\Lambda(\rho)=\sum_{i=1}^2K_i\rho (K_i)^\dagger$ with $K_1=\sqrt{q}|0\ra\la0|+\sqrt{q}|1\ra\la1|$ and $K_2=\sqrt{1-q}(|0\ra\la1|+|1\ra\la0|)$, $0\leq q<1$. 
As for unitary channels, we consider
\begin{equation*}
\begin{gathered}
U=e^{\frac{i\pi \sigma_y}{8}}=
\begin{pmatrix} cos\frac{\pi}{8}  & sin\frac{\pi}{8} \\ -sin\frac{\pi}{8}  & cos\frac{\pi}{8}  \end{pmatrix},~~~
V=e^{\frac{i\pi \sigma_z}{8}}=
\begin{pmatrix} e^{i \frac{\pi}{8}} & 0 \\ 0&e^{-i\frac{\pi}{8}} \end{pmatrix},
\end{gathered}
\end{equation*}
which correspond to the Bloch sphere rotations of $-\pi/4$ about the y axis and z axis, respectively. 
As shown in Fig.~\ref{fig}, our lower bounds are tighter than the ones given in Ref.~\cite{fu2019skew} and Ref.~\cite{zhou2023uncertainty}.

\section{Conclusion}

We have derived several Wigner-Yanase skew information-based uncertainty relations for two quantum channels both in sum and product forms. Especially, we have obtained the near optimal lower bound of the sum-form uncertainty relation for two arbitrary unitary channels. To illustrate the performance of our results, we have presented examples to compare our results with Fu's result in Ref.~\cite{fu2019skew} and Zhou's result in Ref.~\cite{zhou2023uncertainty}, which show that our results are better than the corresponding existing ones.
As uncertainty relations play important roles in many quantum information tasks such as quantum cryptography and quantum metrology, our tighter uncertainty relations may give rise to
better characterization of the information processing.

\bigskip
\noindent{\bf Acknowledgments}\, \,
This work is supported by the National Natural Science Foundation of China (NSFC) under Grants 12075159 and 12171044, Beijing Natural Science Foundation (Grant No. Z190005), Academician Innovation Platform of Hainan Province, and Changsha University of Science and Technology (Grant No. 000303923).

\noindent{\bf Data availability}\, All data generated or analyzed during this study are included in the article.

\noindent{\bf Conflict of interest}\, The authors declare no competing interests.
\bibliography{zhang_wysiref}

\end{document}